\documentclass[a4paper,cite,11pt]{JHEP3}

\usepackage{latexsym}
\usepackage{graphicx}
\usepackage{amsmath}
\usepackage{epsfig}

\newcommand{\lbl}[1]{\label{eq:#1}}
\newcommand{ \rf}[1]{(\ref{eq:#1})}

\newcommand{\be}{\begin{equation}}
\newcommand{\ee}{\end{equation}}
\newcommand{\bea}{\begin{eqnarray}}
\newcommand{\eea}{\end{eqnarray}}
\newcommand{\setl}{\setlength\arraycolsep{2pt}}

\newcommand{\noi}{\noindent}
\newcommand{\nn}{\nonumber}
\newcommand{\ra}{\rightarrow}
\newcommand{\Ra}{\Rightarrow}

\newcommand{\cA}{{\cal A}}

\newcommand{\cM}{{\cal M}}

\newcommand{\cO}{{\cal O}}

\newcommand{\Imm}{\mbox{\rm Im}}
\newcommand{\Ree}{\mbox{\rm Re}}

\newcommand{\MeV}{\mbox{\rm MeV}}
\newcommand{\GeV}{\mbox{\rm GeV}}

\newcommand{\with}{\mbox{\rm with}}

\newcommand{\annd}{\mbox{\rm and}}
\newcommand{\foor}{\mbox{\rm for}}

\newcommand{\als}{\alpha_{\mbox{\rm {\scriptsize s}}}}

\newcommand{\gL}{\frac{1-\gamma_{5}}{2}}
\newcommand{\gR}{\frac{1+\gamma_{5}}{2}}

\newcommand{\E}{\mbox{\rm {\tiny E}}}

\title{Chiral Condensates, $Q_7$ and $Q_8$ Matrix Elements and Large--$N_c$ QCD}

\author{ Samuel FRIOT \\ Centre  de Physique Th{\'e}orique \\UMR 6207 du CNRS et des Universit{\'e}s Aix Marseille 1, Aix Marseille 2 et sud Toulon-Var, affili{\'e}e {\`a} la FRUMAM  \\CNRS-Luminy Case 907 F-13288 Marseille Cedex 9, France \\ \\ E-Mail: \email{friot@cpt.univ-mrs.fr} }
\author{David GREYNAT  \\ Centre  de Physique Th{\'e}orique \\UMR 6207 du CNRS et des Universit{\'e}s Aix Marseille 1, Aix Marseille 2 et sud Toulon-Var, affili{\'e}e {\`a} la FRUMAM  \\CNRS-Luminy Case 907  F-13288 Marseille Cedex 9, France \\ \\ E-Mail: \email{greynat@cpt.univ-mrs.fr}}
\author{ Eduardo de RAFAEL \\ Centre  de Physique Th{\'e}orique \\UMR 6207 du CNRS et des Universit{\'e}s Aix Marseille 1, Aix Marseille 2 et sud Toulon-Var, affili{\'e}e {\`a} la FRUMAM  \\ CNRS-Luminy Case 907  F-13288 Marseille Cedex 9, France \\ \\ Grup de F{\'\i}sica Te{\`o}rica and IFAE\\ Universitat Aut\`onoma de Barcelona\\ 08193 Barcelona, Spain \\ \\ Instituci\'o Catalana de Recerca i Estudis Avan\c{c}ats (ICREA) \\ \\E-Mail: \email{EdeR@cpt.univ-mrs.fr}}

\abstract{ The correlation function of a  $V-A$ current with a  $V+A$ current is discussed within the framework of QCD in the limit of a large number of colours $N_c$. Applications to the evaluation of chiral condensates of dimension six and higher, as well as to the matrix elements of the $Q_7$ and $Q_8$ electroweak penguin operators are discussed. A critical comparison with previous determinations of the same parameters has also been made.}

\begin{document}

\section{Introduction}\lbl{int}

We shall be concerned with the correlation function of a left--handed current
 with a right--handed current
\begin{equation}
	 L^{\mu}(x)=\bar{u}(x)\gamma^{\mu}\gL d(x)\quad\annd\quad
	 	R^{\nu}(0)=\bar{d}(0)\gamma^{\nu}\gR u(0)\,,
\end{equation}
in QCD and in the chiral limit where the light quarks are massless. In this limit, the correlation function in question depends only on one invariant amplitude $\Pi_{LR}(Q^2)$ of the euclidean momentum squared $Q^2=-q^2$, with $q$ the momentum flowing through the two--point function (see Fig.~1):   
\be\lbl{twopf}
2i\int d^4x e^{iq\cdot x} \langle 0\vert \mbox{T}\left\{L^{\mu}(x)\,^{\nu}(0)
\right\}
\vert 0\rangle=\left(q^{\mu}q^{\nu}-q^2
g^{\mu\nu}\right)\Pi_{LR}(Q^2)\;.
\ee

\begin{figure}[h]

\begin{center}
\includegraphics[width=0.4\textwidth]{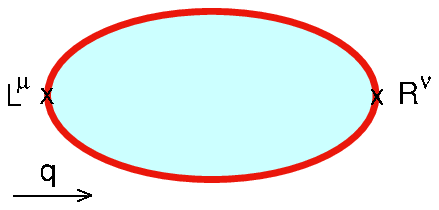}
\end{center}

{\bf Fig.~1} {\it Feynman diagram representing the $\Pi_{LR}$ correlation function in Eq.~\rf{twopf} { in the large-N$_c$ limit}. The solid lines are light quarks propagating in a gluonic background of large--$N_c$ planar diagrams.}

\end{figure}

The interest on the function $\Pi_{LR}(Q^2)$ is twofold: on the one hand, as pointed out in ref.~\cite{KdeR97}, $\Pi_{LR}(Q^2)$ in the chiral limit is an {\it order parameter} of spontaneous chiral symmetry breaking (S$\chi$SB) at all values of $Q^2$; on the other hand, part of the absorptive part of this function i.e.,  $\frac{1}{\pi}\Imm\Pi_{LR}(t)$ with $t=q^2\ge 0$, is accessible to experiment via hadronic $\tau$--decays and $e^+ e^-$ annihilation into hadrons. Furthermore, it has also been shown~\cite{KPdeR98,KPdeR99,DG00,CDG00,KPdeR01} that the same $\Pi_{LR}(Q^2)$ function governs the underlying dynamics of  the leading contributions to the weak matrix elements of the electroweak Penguin--like operators 
\begin{equation}\lbl{4qops}
	Q_7=6(\bar{s}_{L}\gamma^{\mu}d_L)\sum_{q=u,d,s}e_q (\bar{q}_R\gamma_{\mu}q_R)
	\quad\annd\quad Q_8= -12\sum_{q=u,d,s}e_q  (\bar{s}_L q_R)(\bar{q}_R d_L)\,,
\end{equation}
where $e_q$ denote quark electric charges in units of the electric charge and summation over quark color indices within brackets is understood.
These considerations raise the $\Pi_{LR}(Q^2)$ function to the rank of an excellent theoretical laboratory to test new ideas on the fundamental subject of S$\chi$SB in QCD.  

Here, we shall be particularly concerned with the study of $\Pi_{LR}(Q^2)$ in the limit of a large number of colors $N_c$ in QCD. Part of our motivation is to understand the discrepancies between various phenomenological analyses  which have recently been made~\cite{DGHS98,PPdeR01,IZ01,BGP01,CGM03,RL04,Z04} using the experimental data on hadronic $\tau$--decays published by the ALEPH~\cite{aleph} and OPAL~\cite{opal} collaborations at LEP. There are conflicting results for the chiral condensates which modulate the asymptotic behaviour of the $\Pi_{LR}(Q^2)$ function at large $Q^2$ values,    
  between those obtained in refs.~\cite{DGHS98,PPdeR01,IZ01,Z04} and those obtained in refs.~\cite{BGP01,CGM03,RL04}. We want to compare these results to those obtained in two successive approximations to the large--$N_c$ limit: the so{-}called {\it minimal hadronic approximation} (MHA)~\cite{PPdeR01} consisting of a spectrum of a pion state, a vector state and an axial vector state; and the improved approximation where an extra higher  vector state is added. 
  
  In practice, we shall be working with the dimensionless complex function  $W_{LR}[z]$ defined as
\be\lbl{LR}
W_{LR}[{z} ]=-z\Pi_{LR}(z M_{\rho}^2)\,,\quad\with\quad \text{Re}\,z = \frac{Q^2}{M_{\rho}^2}\,,
\ee
and use the mass of the lowest massive state, the $\rho(770~\MeV)$, to normalize quantities with dimensions. In large--$N_c$ QCD the function $ W_{LR}[z]$ is a meromorphic function and, therefore, in full generality, it can be approximated  by  successive  partial fractions of the type 
\be\lbl{LRz}
W_{LR}[z]= A_{N}\prod_{i=1}^{P}\frac{1}{(z+\rho_{i})} \prod_{j=1}^{N}(z+\sigma_{j})\,, \quad\with\quad \rho_1=1 \quad \text{and}\quad \rho_i \neq \rho_k \quad\text{for}\quad i \neq k\;;
\ee
where $P$ (and $N$) get larger and larger, but finite.
On the other hand, in QCD, the operator product expansion (OPE) of the two currents in Eq.~\rf{twopf} fixes the large--$Q^2$ fall off  in $1/Q^2$--powers of the invariant function $\Pi_{LR}(Q^2)$~\cite{SVZ79} to
\be\lbl{OPE}
\lim_{Q^2\ra\infty}\Pi_{LR}(Q^2)=\sum_{n=1}^{\infty}c_{2n+4}(Q^2,\mu^2)
\langle O_{2n+4}(\mu^2)\rangle\frac{1}{(Q^2)^{n+2}}=\frac{1}{2}
\sum_{n=1}^{\infty}
\frac{\langle \cO_{2n+4}\rangle}{(Q^2)^{n+2}} \,.
\ee
Matching the leading asymptotic behaviour for large--$z$ in Eq.~\rf{LRz} to the one of the OPE in Eq.~\rf{OPE}, restricts the number of zeros $N$ and the number of poles $P$ in Eq.~\rf{LRz} to obey the constraint
\be
 N-P=-2\,.
\ee
In the chiral limit, all the vev's $ \langle {\cO}_{2n+4}\rangle$ in  Eq.{\rf{OPE}} are
{\it order parameters} of S$\chi$SB. In
particular, the normalization factor $A_N$ in Eq.~\rf{LRz} is fixed by
the residue of the leading term in the OPE,
\be
A_N=- \frac{1}{2} \frac{1}{(M_{\rho}^2)^{3}}\langle \cO_{6}\rangle\,.
\ee

The case where $N=0$ corresponds to the MHA already mentioned where, besides the Goldstone pole (the pion), which has been removed by the $Q^2$ factor in the r.h.s. of Eq.~\rf{LR}, there are two poles (since $P=2$ in this case): the lowest vector state and an axial--vector state~\cite{KdeR97}.

Strictly speaking, the $\langle \cO_{2n+4}\rangle$ in the r.h.s. of Eq.~\rf{OPE} are $Q^2$ dependent, because of the $\log Q^2$ dependence of the Wilson coefficients $c_{2n+4}(Q^2,\mu^2)$ via the pQCD series in $\als$. This $\log Q^2$ dependence can only be reproduced if $P\ra\infty$ (and $N\ra\infty$) in Eq.~\rf{LRz}. In writing the large--$N_c$ approximation to the $W_{ LR}[z]$ function in Eq.~\rf{LRz}, one is implicitly assuming an effective cancellation between the extra poles and zeros in the complex $z$--plane which lye beyond a disc of radius $s_0$ covering all the poles and zeros retained in that approximation. In the Minkowski axis, this is equivalent to assuming that, for all practible purposes, there is a cancellation between the $V$--spectral function and the $A$--spectral function for $t\geq s_0$, an assumption which is consistent with the fact that in pQCD and in the chiral limit these two spectral functions are identical. The same approximation, in the deep euclidean region, suggests that when comparing the asymptotic inverse powers of $z$  in Eq.~\rf{LRz} to those of the OPE in Eq.~\rf{OPE}, the $\log Q^2$ dependence of the Wilson coefficients in $c_{2n+4}(Q^2,\mu^2)$ be frozen at $Q^2\sim s_0$.
  
\section{General Properties of $\Pi_{LR}(Q^2)$ in Large--$N_c$ QCD}
\setcounter{equation}{0}

The function $\Pi_{LR}(Q^2)$ is the simplest case of a class of Green's functions, which we call generalized two--point functions, and  which obey rather remarkable {\it short--distance
$\rightleftharpoons$ long--distance duality} properties. Generalized two--point functions are two--point functions with a finite number of local operator insertions carrying zero momenta, (no insertions in the case of $\Pi_{LR}(Q^2)$). These Green's functions, in the large--$N_c$ limit, are meromorphic functions which can be approximated by polynomial ratios
\be\lbl{shortlong}
W[z;\ \rho_{1},\rho_{2},\cdots\rho_{P};\ \sigma_{1},\sigma_{2}\cdots
\sigma_{N}]=A_{N}\prod_{i=1}^{P}\frac{1}{(z+\rho_{i})}
\prod_{j=1}^{N}(z+\sigma_{j})\,;
\ee
and they are fully determined by their poles, their zeros and the overall normalization $A_N$.
Using this representation, one can readily see that
\be
 W\left[\frac{1}{z};\ \frac{1}{\rho_{1}},\frac{1}{\rho_{2}}\cdots
\frac{1}{\rho_{P}};\ \frac{1}{\sigma_{1}},\frac{1}{\sigma_{2}}\cdots
\frac{1}{\sigma_{N}}\right]=z^{P-N}\ \frac{\prod \rho_{i}}
{\prod\sigma_{j}}\ W[z;\ \rho_{1},\rho_{2},\cdots\rho_{P};\ 
\sigma_{1},\sigma_{2}\cdots\sigma_{N}]\,,
\ee
relating the OPE expansion (powers of $1/z$) to the chiral expansion (powers of $z$).
In the case of the $W_{LR}[z]$ correlation function in Eq.~\rf{LR} this is the  generalization to an arbitrary number of narrow states of
the simple relation~\cite{deR99}
\begin{equation}
W_{ LR}\left[\frac{1}{z};\ \frac{1}{\rho_{1}},\frac{1}{\rho_{2}}\right]=z^{2}\ \rho_1 \rho_2 W_{ LR}[z;\ \rho_{1},\rho_{2}]\,,\quad\foor\quad N=0\,,	
\end{equation}
corresponding to the case of the MHA, where $P=2$ and $N=0$. 

One can easily deduce the following set of properties, useful for phenomenological applications:

\begin{enumerate}
\item {\sc The $W_{ LR}[z]$ function at the origin}
\be
W_{LR}[0]=A_{N}\frac{\sigma_{1}\sigma_{2}\dots\sigma_{N}}
{\rho_{1}\rho_{2}\dots\rho_{P}}\quad\foor\quad N\ge 1
\quad\annd\quad W_{LR}[0]=\frac{A_{0}}{\rho_{1}\rho_{2}}
\quad\foor\quad N=0\quad(\mbox{\rm MHA})\,.
\ee
Recall that the value of $W_{LR}[0]$ is fixed by lowest order chiral
perturbation theory; so, once $A_N$ is known, {\it the sign of $W_{LR}[0]$ fixes the sign
of the product of all the zeros}~\footnote{Notice that, if there are  complex zeros, they have
to appear in  conjugate pairs of each other and therefore give a
positive contribution to the product.}. We therefore have that

\be
A_{N}\frac{\sigma_{1}\sigma_{2}\dots\sigma_{N}}
{\rho_{1}\rho_{2}\dots\rho_{P}}=\frac{F_{0}^2}{M_V^2}  \equiv \rho_F \quad\foor\quad
N\ge 1
\quad\annd\quad \frac{A_{0}}{\rho_{1}\rho_{2}}=\frac{F_{0}^2}{M_V^2}
\quad\foor\quad N=0\,,
\ee 
with $F_0^2$ the residue of the Goldstone pole (the pion) contributing to $\Pi_{LR}(Q^2)$.

In what follows, we shall often use the notation $\rho_2 \equiv \rho_A$ and $\rho_3 \equiv \rho_{V'}$.

\item{ \sc The $W_{LR}[z]$ function at infinity: condensates}

Starting from equation~\rf{LRz} with the constraint $N-P=-2$ (which means that the two Weinberg sum rules are automatically satisfied) and decomposing it into partial fractions, we obtain 
\begin{equation}
\label{Wdecompo}
 W_{LR}[z]= A_{P-2}\prod_{i=1}^{P}\frac{1}{(z+\rho_{i})} \prod_{j=1}^{P-2}(z+\sigma_{j}) = \sum_{k=1}^{P} w_k \frac{1}{z+\rho_k}\;,
\end{equation}
with 
\begin{equation}
\label{wk}
 w_k = A_{P-2} \prod_{i\neq k = 1}^{P} \frac{1}{\left(\rho_i - \rho_k\right)} \prod_{j=1}^{P-2}\left(\sigma_j-\rho_k\right) = \rho_F \prod_{i= 1}^{P} \frac{\rho_i}{\left(\rho_i - \rho_k + \delta_{ik}\right)} \prod_{j=1}^{P-2}\frac{\left(\sigma_j-\rho_k\right)}{\sigma_j}\;,
\end{equation}
where $\delta_{ik}$ is the Kronecker symbol. This fixes the residues of the poles, which are physical couplings, in terms of the poles and zeros.
 There follows then that:  
\begin{equation}
\label{condensatgene}
\left < \cO_{2n+4} \right> = 2 (-1)^{n+1}\, M_V^{2n+4}\, \sum_{k=1}^P  w_k \rho_k^n  \;. 
\end{equation}

\item{\sc The Linear Constraint} 

This is a very interesting constraint,  which simply
follows by expanding Eq.~\rf{LRz} to first non--trivial order in
inverse powers of
$z$
\be\lbl{linear}
\sum_{j=1}^{N}\sigma_{j}-\sum_{i=1}^{P}\rho_{i}=\frac{1}{M_V^2}\frac{\langle
\cO_{8}\rangle}{\langle\cO_{6}\rangle}\,.
\ee
It relates the difference between  the sum of the
positions of the zeros and the sum of the  positions of the poles to the ratio of next--to--leading to leading
vev's in the OPE. In other words, once we know the positions of the
poles, the sum of the positions of the zeros, which is always real, is governed by the
next--to--leading order term in the OPE. In the case corresponding to
the MHA, where by definition there are no zeros, this constraint
simply becomes:
\be\lbl{O8O6}
1+\rho_{A}=-\frac{1}{M_V^2}\frac{\langle
\cO_{8}\rangle}{\langle\cO_{6}\rangle}\,,
\ee
implying that, in the MHA  $\langle
\cO_{8}\rangle$ and $\langle
\cO_{6}\rangle$ must have {\it opposite signs}. This is indeed what the phenomenological analyses in refs.~\cite{DGHS98,PPdeR01,IZ01,Z04} find, in contradistinction to the results in refs.~\cite{BGP01,CGM03,RL04} which find the same sign for the first two condensates.
In fact, in the case of the MHA, we can show that this alternance of sign for two consecutive condensates is a generic property. It follows from the relation
\begin{equation}
\label{OnMHA}
\left < \cO_{2n+4} \right> = 2 (-1)^{n}\, M_V^{2n+4}\, A_0 \,\frac{1-\rho_A^n}{1-\rho_A} = 2 (-1)^{n}\,  M_V^{2n+4}\, \rho_F\, \sum_{k=1}^{n}\rho_A^k\,, \quad\foor\quad n\ge 1\,.
\end{equation}
Except for the $(-1)^n$ factor, all the quantities appearing in the r.h.s. are positive; which explains the alternance in sign with $n$.

\vspace*{0.25cm}\item{\sc The Slope Constraint}

The value of the derivative of $W_{LR}[z]$ at $z=0$ is controlled by
the $\cO(p^4)$ $\chi$PT low energy constant $L_{10}$~\cite{GL85} 
\be
W_{LR}'[0]=4 L_{10}\,.
\ee
The relation to the poles and zeros of the $W_{LR}[z]$--function in large--$N_c$ QCD is
as follows
\be
W_{LR}'[0]=\rho_F \left ( \sum _{j=1}^N \frac{1}{\sigma_j} - \sum_{i=1}^P \frac{1}{\rho_i} \right) \,,\quad\foor\quad N\ge 1\;.
\ee

In the case $N=1$ it reduces to the interesting relation
\be
4L_{10}=\frac{F_{0}^2}{M_{V}^2}
\left[\frac{1}{\sigma}-
\left(1+\frac{1}{\rho_{A}}+\frac{1}{\rho_{V'}}\right)\right]
\,,
\ee
and  in the case $N=0$, corresponding to the MHA, to the well--known
result~\cite{EGPdeR89,EGLPdeR89}
\be
4L_{10}=-\frac{F_{0}^2}{M_{V}^2} 
\left(1+\frac{1}{\rho_{A}}\right)
\,\quad\Ra\quad
L_{10}=-\frac{3}{8}\frac{F_{0}^2}{M_{V}^2}\quad\foor\quad \rho_A=2 \,,  
\ee
indicating that in the MHA, the slope of $W_{LR}[z]$ at the origin has the
opposite sign to the value of $W_{ LR}[0]$, in agreement with experiment.

\vspace*{0.25cm}
\item{\sc Dispersion Relations}

The function $W_{LR}[z]$ and its corresponding spectral function  $\frac{1}{\pi}\Imm \Pi_{LR}(t)$ are related to each other by the dispersion relation
\begin{equation}
\label{disp}
W_{LR}\left(\text{Re}\,z\right)=-\frac{Q^2}{M_{\rho}^2}\int_{0}^{\infty}dt\frac{1}{t+Q^2 -i\epsilon}\frac{1}{\pi}\Imm \Pi_{LR}(t)\,.
\end{equation}
This is the {\it Hilbert transform} of a spectral function, which in our approximation of large--$N_c$ QCD has the general form
\begin{equation}
\label{spectre}
	\frac{1}{\pi}\Imm \Pi_{LR}(t)=-F_0^2 \delta(t)+\sum_{i=1}^{P}\alpha_{i} M_{i}^2 \delta(t-M_{i}^2)\,
\end{equation}
with the sum ordered in increasing values of the masses $M_{i}^2$.
The residues $\alpha_{i}$ are positive for the vector states and  negative for the pion pole (the first term) and the  axial states.

Sometimes it is also convenient to consider the {\it Laplace transform (Borel transform)} of the spectral function
\begin{equation}
	\cM_{LR}(\sigma)=\frac{1}{M_{\rho}^2}\int_{0}^{\infty}dt e^{-t\sigma}\frac{1}{\pi}\Imm\Pi_{LR}(t)\,.
\end{equation}
The two types of transforms are related by the fact that
\begin{equation}
	\frac{1}{t+Q^2}=\int_{0}^{\infty}d\sigma e^{-t\sigma}e^{-Q^2\sigma}\,;
\end{equation}
therefore,
\begin{equation}
W_{LR}(z)=	-zM_{\rho}^2\int_{0}^{\infty}d\sigma e^{-\sigma  M_{\rho}^2 z}	\cM_{LR}(\sigma)\,.
\end{equation}

In practice, one is often interested in observables which are  moments of $W_{LR}(Q^2)$
\begin{equation}
	\cO^{(m)}=\int_{0}^{\infty}dz\ z^{m}\  W_{LR}(z)=-\frac{1}{\left(M_{\rho}^{2}\right)^{1+m}}\int_{0}^{\infty}\frac{d\sigma}{\sigma^{2+m}}\cM_{LR}(\sigma)\,;\quad m=0,1,2,\cdots\,,
\end{equation}
showing that we can view the observables $\cO^{(m)}$, either as moments of $W_{LR}(z)$, or as inverse moments  of  $\cM_{LR}(\sigma)$.

Some interesting properties of the Laplace Transform follow.

\begin{itemize}
\item The Laplace transform $\cM_{LR}[\sigma]$ obeys the differential
equation
\be\lbl{difeq}
\left\{a_P\frac{d^P}{d\sigma^P}+a_{P-1}\frac{d^{P-1}}{d\sigma^{P-1}}+\cdots
+ a_{1}\frac{d}{d\sigma}+a_0\right\}\cM_{LR}[\sigma]=0\,,
\ee
with $a_P$, $a_{P-1}$, ... $a_0$ the coefficients of the polynomial
\be
\prod_{i=1}^{P}(z+\rho_i)=a_{P}z^P+a_{P-1}z^{P-1}+\cdots +a_0\,;
\ee
i.e.,
\begin{equation}
	a_P=1\,,\quad a_{P-1}=\sum_{i=1}^{P}\rho_i\,,\quad \cdots\quad a_0=\prod_{i=1}^{P}\rho_i\,.
\end{equation}

\item The polynomial $\prod_{i=1}^{P}(z+\rho_i)$ is a {\it stable
polynomial}, because all its roots are in the
negative real axis (the Minkowski axis).

\item
The most general solution of the differential equation in \rf{difeq}
is of the form
\be
-\frac{F_{0}^2}{M_{\rho}^2}+\alpha_{1}\rho_{1} e^{-\rho_1 \sigma}+ \alpha_{2}\rho_{2} e^{-\rho_2 \sigma}+\cdots
+\alpha_{P}\rho_{P} e^{-\rho_P \sigma}\,,
\ee
with the $\alpha_{i}$ constants fixed by the successive boundary
conditions obtained e.g., from the knowledge of the derivatives of the
function $\cM_{LR}[\sigma]$ at the origin.This leads to
the system of equations first discussed in ref.~\cite{KdeR97}:

\begin{equation}\lbl{genwein}
\left\{	\begin{array}{lllllllll}
	& -\frac{F_{0}^2}{M_{\rho}^2}  + &\alpha_{1}\rho_1 & + &\alpha_{2}\rho_2 &+\cdots+ &\alpha_{P}\rho_P & = & \cM_{LR}[0]\,, \\ & & 
\alpha_{1}\rho_{1}^2 &+&\alpha_{2}\rho_{2}^2&+\cdots+&\alpha_{P}\rho_{P}^2 & = &
\frac{d \cM_{LR}[0]}{d(\sigma M_{\rho}^2)}\,, \\  & & & & & & &\cdots & \\ &  
&\alpha_{1}\rho_{1}^{P+1}&+&\alpha_{2}\rho_{2}^{P+1}& +\cdots+&\alpha_{P}\rho_{P}^{P+1} &
= &
\frac{d^{P} \cM_{LR}[0]}{d(\sigma M_{\rho}^2)^P}\,.
	\end{array}\right.
\end{equation}

\noi

The discriminant of this system of equations in the $\alpha_i$ is the Vandermonde
determinant
\be
\left(
\begin{array}{cccc}
1 & 1 & \cdots & 1 \nn \\
\rho_1 & \rho_2 & \cdots & \rho_P \nn \\
 & & \cdots  & \nn \\
\rho_{1}^{P} & \rho_{2}^{P} & \cdots & \rho_{P}^{P}
\end{array}
\right)\,.
\ee
On the other hand the successive values of $\cM_{LR}[0]$, $\frac{d \cM_{ LR}[0]}{d\sigma}$,
$\cdots$ and $\frac{d^{P} \cM_{LR}[0]}{(d\sigma)^P}$ are fixed by the OPE
of the function $W_{LR}[z]$. In our case $\cM_{LR}[0]=0$; $\frac{d \cM_{ LR}[0]}{d\sigma}=0$ and the first two equations in \rf{genwein} are nothing but the well--known  1st and 2nd Weinberg sum rules.

\item
The positive moments of the Laplace transform, with the Goldstone singularity removed, correspond to coefficients of the chiral expansion
\begin{equation}
	\int_0^{\infty} d\sigma \sigma^N \cM_{LR}(\sigma)=	\Gamma(N+1)\int_0^{\infty} dt\frac{1}{t^{N+1}}\frac{1}{\pi}\Imm {\tilde \Pi}_{LR}(t)\,,\quad\foor\quad N\ge 0\,,
\end{equation}
where
\begin{equation}
\frac{1}{\pi}\Imm {\tilde \Pi}_{LR}(t)=\frac{1}{\pi}\Imm  \Pi_{LR}(t)	+F_{0}^2 \delta(t)\,.
\end{equation}

\end{itemize}

\item{ \sc  Reconstruction of the Spectral Function}

 It is useful to give the coefficients $\alpha_i$ of equation (\ref{spectre}) in terms of the poles $\rho_i$ and zeros $\sigma_j$. 

Inserting (\ref{spectre}) into (\ref{disp}) we find 
\begin{equation}
\label{Wspectre}
 W_{LR}[z]= \rho_F - \sum_{k=1}^P \alpha_k \rho_k + \sum_{k=1}^P \alpha_k \rho_k^2 \frac{1}{z+\rho_k}\;.
\end{equation}

By comparing (\ref{Wdecompo}) and (\ref{Wspectre}), we then have 
\begin{equation}
\label{ident}
 \begin{cases}
&\displaystyle{\rho_F - \sum_{k=1}^P \alpha_k \rho_k=0_,, } \\
&\\
&\alpha_k \rho_k^2 = w_k\;, \quad\with\quad \sum_{k=1}^{P} w_k =0\,,
\end{cases}
\end{equation}
where the two sum equations above are just the first and second Weinberg sum rules. 

Using a more conventional notation in terms of axial to vacuum and vector to vacuum couplings: $\alpha_A = - f_A^2$ and $\alpha_V = f_V^2$, we conclude from equations (\ref{ident}) and (\ref{wk}) that 

\begin{align}
&f_V^2 M_V^2 =  \frac{F_0^2}{\rho_V} \prod_{i= 1}^{P} \frac{\rho_i}{\left(\rho_i - \rho_V + \delta_{iV}\right)} \prod_{j=1}^{P-2}\frac{\left(\sigma_j-\rho_V\right)}{\sigma_j} \label{fV}\\
& \nonumber\\
&f_A^2 M_A^2 = - \frac{F_0^2}{\rho_A} \prod_{i= 1}^{P} \frac{\rho_i}{\left(\rho_i - \rho_A + \delta_{iA}\right)} \prod_{j=1}^{P-2}\frac{\left(\sigma_j-\rho_A\right)}{\sigma_j}\label{fA}\;.
\end{align}

\vspace*{0.25cm}
\item{\sc The Smoothness Assumption}

As already stated, the MHA corresponds to the case where $N=0$. In this approximation, the function $W[z]$, or its equivalent  $\cM(\sigma)$, is a monotonous  function of the euclidean variable: $0\le Q^2 \le \infty$ or, equivalently, $\infty\ge \sigma \ge 0$. Introducing an improved approximation with more poles necessarily brings in non--trivial zeros. Can the zeros change dramatically the smoothness of the MHA? So far, all the calculations made in the literature in the MHA are based on the working assumption that the smoothness, beyond the MHA, persists; in other words, one is assuming an underlying {\it hypothesis of smoothness} of large--$N_c$ QCD  which, in full generality, has not been  proved from first principles. What follows in the next sections is a test of this assumption, albeit in a very particular case.
 
We can only suggest a possible scenario on how this {\it smoothness assumption} may be a generic property of large--$N_c$ QCD.    
It is quite clear that in the case where the spectral function is positive, the fact that $\cM(\sigma)$ is then {\it logarithmically convex}~\cite{deR98} provides the required smoothness property; however, in most cases (like the left--right correlation function we are considering here) this property of positivity does not hold. 

In fact, in the case of the left--right correlation function, Witten has proved~\cite{W83}, under rather general assumptions, that $\Pi_{LR}(Q^2)\ge 0$ for $0\le Q^2 \le \infty$. This follows from the positivity of the measure in the gluonic path integral and Schwartz type inequalities of the fermion propagators, Witten's proof, however, is not in general applicable to other generalized two--point functions. 

A property of smoothness would follow if one could guarantee that  the positions of the  zeros of the generalized Green's functions are all in the negative half--plane $\text{Re}\,z\le 0$ (i.e. the half--plane which includes the Minkowski axis). This would be  the case if the polynomial of  zeros $\prod_{j=1}^{N}(z+\sigma_{j})$, like the polynomial of poles $\prod_{i=1}^{P}(z+\rho_{i})$, 
was also a stable polynomial~\footnote{ In the mathematical literature this goes under the name of the {\it Routh--Hurwitz theorem}. See e.g. ref.~\cite{stable}. We have recently been able to prove that the successive polynomials $\prod_{j=1}^{N}(z+\sigma_{j})$ for $N=1,2,\cdots$ up to an arbitrary, but finite $N$, are indeed stable polynomials. The proof, however, lies outside the context of this paper and will be published elsewhere.}. 

\end{enumerate}

\section{$\Pi_{LR}$ in the case of a $\pi-V-A-V'$ Spectrum}
\setcounter{equation}{0}

It is claimed by some of the authors of refs.~\cite{BGP01,CGM03,RL04} that the reason why  their phenomenological
analysis of the chiral condensates  give the same sign for $\langle\cO_6\rangle$ and
$\langle\cO_8\rangle$ is due to the fact that the hadronic $\tau$--decay
spectrum is sensitive to the presence of the $\rho'$, while the MHA
ignores all higher states beyond the first axial state. Partly motivated by this claim~\footnote{Notice, however, that there  are  other phenomenological analyses which using the same $\tau$--data find opposite signs for $\langle\cO_6\rangle$ and
$\langle\cO_8\rangle$~\cite{DGHS98,IZ01,Z04}}, we want to analyze here the case, beyond the MHA, where an extra vector state $V'$, and therefore one zero $\sigma$, are also included. Let us collect the relevant equations
corresponding  to this case.

\begin{enumerate}
\vspace*{0.25cm}\item{\sc The Correlation Function}

With a spectrum of a pion pole, and $V$, $A$, and $V'$ states, the relevant correlation function is
\be\lbl{LRV2}
-\frac{Q^2}{M_V^2}\Pi_{LR}(Q^2)\Longrightarrow W_{LR}[z]=A_1
\frac{z+\sigma}{(z+1)(z+\rho_A)(z+\rho_{V'})}\,,
\ee
where
\be
A_1\frac{\sigma}{\rho_A \rho_{V'}}=\frac{F_0^2}{M_V^2}\equiv \rho_F\,\quad\annd\quad
\langle\cO_6\rangle=\frac{-2}{\sigma}F_0^2 M_A^2 M_{V'}^2=-M_V^6 \frac{2}{\sigma} \rho_F \rho_A \rho_{V'}\,.
\ee
Because of the positivity of $W_{LR} [z]$ for $\Ree\,z\ge 0$, the position of the zero
has to be in the Minkowski axis and, therefore, $\sigma>0$. 

\vspace*{0.25cm}\item{\sc The Linear Constraint}

Equation \rf{linear} now reduces to the simple relation
\be
\sigma-(1+\rho_A+\rho_{V'})=\frac{1}{M_V^2}\frac{\langle\cO_8\rangle}
{\langle\cO_6\rangle}\,.
\ee
This is one of the key equations of our discussion, which already provides a semiquantitative argument in favor of the opposite sign option for the condensates $\langle\cO_{ 6} \rangle$ and $\langle\cO_8\rangle$ . The equation  states that for
$\langle\cO_8\rangle$ to have the same sign as $\langle\cO_6\rangle$,
the position of the zero has to be {\it far beyond the largest
$V'$--pole}:
\be\lbl{zp}
\sigma > 1+\rho_A+\rho_{V'}\,.
\ee
Fixing the position of the poles at the values of the observed spectrum
(and ignoring errors for the purpose of the discussion){,} one has
\be
M_V={0.776}~\GeV\,,\quad
M_A=1.230~\GeV\ (\rho_{A}=2.5)\,,\quad
M_{V'}=1.465~\GeV\ (\rho_{V'}=3.6)\,;
\ee
which means that for the equal sign requirement option to be satisfied one must have $\sigma>7.2$. In
$\GeV$ units this corresponds to a mass of $2.1~\GeV$. Now,
as already stated at the end of section I, in writing a large--$N_c$ ansatz for the $W_{LR}[z]$ function, one is implicitly assuming an effective cancellation between the extra poles and zeros in the complex $z$--plane which lye beyond a disc of radius $s_0$ covering all the poles and zeros retained in that approximation. The result $\sigma>7.2$
implies that the radius in question has to be $\sqrt{s_{0}}>2.1~\GeV$. A
priori that  {\it seems a good thing} because the OPE--matching is now applied at $Q^2\ge s_0$; i.e. in a
more asymptotic region than in the case of the MHA ansatz; however, it
also implies that there are no further poles in the region between  
$M_{V'}\simeq 1.5$ and the effective mass $M_{\sigma}\simeq 2.1~\GeV$ corresponding to the zero at $\sigma\simeq 7.2$. This, however,  is in contradiction
with the observed $a_1$--like state at $M_{A'}\simeq 1.64 ~\GeV$ and $\rho$--like states at $M_{V''}\simeq 1.72~\GeV$ and $M_{V'''}\simeq 1.9~\GeV$ below $M_{\sigma}\simeq 2.1~\GeV$. Alternatively, if one excludes those three states  $A'$, $V''$ and $V'''$ as all the phenomenological analyses using $\tau$--data do in fact, then the position of the zero $\sigma$ should be $\sigma\lesssim \frac{M_{A'}^2}{M_V^2}\approx4.5$, implying according to eq.~\rf{zp}, that $\langle \cO_8\rangle$ and $\langle \cO_6\rangle$ must have opposite signs, in contradiction with the claims of refs.~\cite{BGP01,CGM03,RL04}.

\vspace*{0.25cm}\item{\sc The Slope Constraint}

This is the relation between $L_{10}$ and the position of the poles and the zero
\be\lbl{L10}
4L_{10}=\rho_F
\left[\frac{1}{\sigma}-
\left(1+\frac{1}{\rho_{A}}+\frac{1}{\rho_{V'}}\right)\right]
\,,
\ee
which was already discussed in the previous section. Since $L_{10}$ is relatively well--known phenomenologically it gives a constraint between $\rho_F$, $\rho_A$, $\rho_{V'}$ and $\sigma$.

\vspace*{0.25cm}\item{\sc The Electromagnetic $\pi^{+}-\pi^{0}$ Mass Difference}

Recall that
\be
m_{\pi^+}^2\vert_{\mbox{\tiny\rm
em}}=\frac{3}{4}\frac{\alpha}{\pi}\frac{1}{F_0^2}
\int_{0}^{\infty} dQ^2 \left[-Q^2\Pi_{LR}(Q^2)\right]\,.
\ee
In the MHA with a $V$--$A$ spectrum,
\be
m_{\pi^+}^2\big\vert_{{\mbox\tiny\rm em}}^{\mbox{\tiny\rm (MHA)}}=
\frac{3}{4}\frac{\alpha}{\pi}M_{V}^2\rho_{A}\frac{1}{\rho_{A}-1}
\log{\rho_{A}}\,.
\ee
In the case of a  V--A--V' spectrum we find
\be\lbl{emd}
m_{\pi^+}^2\big\vert_{{\mbox\tiny\rm em}}^{\mbox{\tiny\rm (V--A--V')}}=
\frac{3}{4}\frac{\alpha}{\pi}M_{V}^2\rho_{A}\frac{\rho_{V'}}{\sigma}
\frac{(\rho_{V'}-1)(\rho_{A}-\sigma)\log\frac{\rho_{V'}}{\rho_{A}}+
(\rho_{V'}-\rho_{A})(\sigma-1)\log{\rho_{V'}}}
{(\rho_{V'}-1)(\rho_{A}-1)(\rho_{V'}-\rho_{A})}\,,
\ee
which, for $\sigma=\rho_{V'}$, reduces to the MHA expression. Since the $\Delta m=m_{\pi^+}-m_{\pi^0}$ mass difference is dominated by its electromagnetic contribution, we can use its experimental value  as a further constraint on $\rho_A$, $\rho_{V'}$ and $\sigma$.

\vspace*{0.25cm}

\item{\sc Reconstruction of the Spectral Function}

In full generality, as shown in the previous section, one can reconstruct the spectral function from the knowledge of the zeros and poles and the normalization $A_N$. In our case, using (\ref{fV}) and (\ref{fA}), this results in
\be
\frac{1}{\pi}\Imm\Pi_{LR}(t)=
\frac{1}{\pi}\Imm\Pi_{V}(t)-\frac{1}{\pi}\Imm\Pi_{A}(t)\,,
\ee
with
\be
\frac{1}{\pi}\Imm\Pi_{A}(t)=F_{0}^2\delta(t)+
F_{0}^2
\frac{\rho_{V'}}{\sigma}
\frac{\sigma-\rho_{A}}{(\rho_{A}-1)(\rho_{V'}-\rho_A)}
\delta(t-M_A^2) + \frac{N_c}{16\pi^2}\frac{2}{3} \theta(t-s_0)(1+\cdots)\,,
\ee
and
\begin{align}
\frac{1}{\pi}\Imm\Pi_{V}(t)\!&=\!
F_{0}^2
\frac{\rho_{A}
}{\sigma}\left\{
\frac{\rho_{V'}(\sigma-1)}{(\rho_{A}-1)(\rho_{V'}-1)}
\delta(t\!-\!M_{V}^2)+
\frac{\sigma-\rho_{V'}}{(\rho_{V'}-1)(\rho_{V'}-\rho_A)}
\delta(t\!-\!M_{V'}^2)\right\} \nonumber \\
&\hspace{4cm}+ \frac{N_c}{16\pi^2}\frac{2}{3} \theta(t-s_0)(1+\cdots)\,,
\end{align}
where the dots in these equations stand for pQCD $\als$--corrections.
Notice that these spectral functions, by construction, satisfy the 1st and 2nd Weinberg sum rules. For $\sigma=\rho_{V'}$ they reduce to
the spectral functions of the MHA case.

\vspace*{0.25cm}\item{\sc Radiative Widths}

For a meson V in the lowest octet of vector states, the width of the electronic decay $V\ra e^{+}
e^{-}$ is given by the expression 
\be\lbl{rhow}
\Gamma_{V \ra e^{+}e^{-}}=\frac{4\pi\alpha^2}{3}f_{V}^2 M_{V}\,.
\ee

For an axial--vector A, the width of the decay $A\ra\pi\gamma$, in the chiral limit, is
given by~\cite{EGPdeR89}

\be\lbl{Aw}
\Gamma_{A\ra\pi\gamma}=\frac{\alpha}{24}f_{A}^2
\frac{M_{A}^2}{F_{0}^2}M_{A}
\,.
\ee
The relation between these decay rates and $\rho_A$, $\rho_{V'}$ and $\sigma$ is as follows:

{\setl
\bea
f_{V}^2 M_V^2 & = & F_{0}^2
\frac{\rho_{A}\rho_{V'}}{\sigma}
\frac{\sigma-1}{(\rho_{A}-1)(\rho_{V'}-1)}\,, \\
f_{A}^2 M_A^2 & = & F_{0}^2
\frac{\rho_{V'}}{\sigma}
\frac{\sigma-\rho_{A}}{(\rho_{A}-1)(\rho_{V'}-\rho_A)}\,, \\
f_{V'}^2 M_{V'}^2 & = & F_{0}^2
\frac{\rho_{A}}{\sigma} \frac{\sigma-\rho_{V'}}{(\rho_{V'}-1)(\rho_{V'}-\rho_A)}\,.
\eea}

\vspace*{0.25cm}
\item{\sc The Electromagnetic Pion form Factor}
 
In Large--$N_c$ QCD, the electromagnetic form factor of the pion has a particularly simple expression
\begin{equation}
	F(t)=1+\sum_{V}\frac{F_{V}G_{V}}{F_{0}^2}\frac{t}{M_{V}^2-t}\,,
\end{equation}
where $F_V=f_{V}M_{V}$ and $G_V=g_{V}M_{V}$ are standard couplings of the large--$N_c$ effective  Lagrangian of narrow states~\cite{EGPdeR89}. 
Requiring that the form factor falls as an inverse power of $Q^2=-t$ in the deep euclidean, one gets the constraint:
\begin{equation}\lbl{norinf}
	\sum_{V} f_{V}g_{V}M_{V}^2=F_{0}^2\,.
\end{equation}
On the other hand the slope at the origin of the pion electromagnetic form factor determines the $L_9$ coupling~\cite{GL85} as follows:
\begin{equation}\lbl{l9con}
	L_9 =\frac{1}{2}\sum_{V} f_{V}g_{V}\,.
\end{equation}

The two constraints \rf{norinf} and \rf{l9con}, when restricted to a $\pi$, $V$, $A$, $V'$ spectrum, become:

{\setl
\bea
\label{eq1}f_{V}g_{V}+f_{V'}g_{V'}\rho_{V'} & = & \rho_{ F}\,, \\
\label{eq2}f_{V}g_{V}+f_{V'}g_{V'} & = & 2L_9\,.
\eea}

\noi
In the chiral limit, the coupling $g_V$ is related to the $\rho\ra\pi\pi$ width as follows
\begin{equation}
	\Gamma_{\rho\ra\pi\pi}=g_{V}^2\frac{M_{\rho}^5}{48\pi F_{0}^4}\,.
\end{equation}
The $V'$--width, however, (and hence the coupling $g_{V'}$) is poorly known. We can eliminate $g_{V'}$  between the two equations (\ref{eq1}) and (\ref{eq2}), which results in a {\it useful} constraint between $L_9$, $\rho_A$, $\rho_{V'}$, $\sigma$ and $g_V$ (which can be fixed from $\rho\ra\pi\pi$). 

\vspace*{0.25cm}
\item{\sc Matrix Elements of the $Q_7$ and $Q_8$ Operators}

The four quark operators in question are the ones in eq.~\rf{4qops}. We are interested in the evaluation of the  matrix elements of these operators between an incoming $K$--state and an outgoing $2\pi$--state. 

\begin{description}

\vspace*{0.25cm}
\item[Evaluation of $\langle O_1 \rangle$ ]

As discussed in ref.~\cite{KPdeR01}, the
$\langle(\pi\pi)_{I=2}\vert Q_7\vert K^0 \rangle$ matrix element, to lowest order in $\chi$PT is related to the vev 
\begin{equation}
	\langle O_1\rangle \equiv \langle 0\vert (\bar s_L \gamma^{\mu}d_L) (\bar d_R \gamma_{\mu}s_R)\vert 0\rangle\,,
\end{equation}
as follows
\be
\lbl{Q7}
M_7 \equiv
\langle(\pi\pi)_{I=2}\vert Q_7\vert K^0 \rangle = -\frac{4}{F_0^3}\langle O_1\rangle\,.
\ee
On the other hand, in ref.~\cite{KPdeR01}, it was also shown that 
\begin{equation}
\langle O_1 \rangle=\frac{1}{6}\left(-3ig_{\mu\nu}\int\frac{d^4 q}{(2\pi)^4}\Pi_{LR}^{\mu\nu}(q)\right)_{\overline{\mbox{\rm\tiny MS}}}\,,
\end{equation}
where the integral has to be evaluated using the same $\overline{\mbox{\rm MS}}$--renormalization prescription as used for the evaluation of the corresponding Wilson coefficient of $Q_7$.  One then  finds that  in large--$N_c$ QCD~\cite{KPdeR01} 
\begin{equation}
	\langle O_1\rangle  = -\frac{3}{32\pi^2}\left[\sum_A f_A^2 M_A^6 \log\frac{\Lambda^2}{M_A^2}-\sum_V f_V^2 M_V^6 \log\frac{\Lambda^2}{M_V^2}\right]\,,
\end{equation}
where $\Lambda^2=\mu^2 \exp(1/3+\kappa)$ with $\kappa$ depending on the renormalization scheme:  $\kappa=-1/2$ in NDR and $\kappa=+3/2$ in HV. For a $\pi$--$V$--$A$--$V'$ spectrum this can be written as a function of $\rho_A$, $\rho_{V'}$ and $\sigma$ in the following way 

{\setl
\bea
\langle O_1\rangle 
 & = & \frac{3 M_V^6}{32\pi^2} \rho_F\frac{\rho_{V'}}{\sigma}\left[\rho_{A}\frac{\sigma-1}{(\rho_A-1)(\rho_{V'}-1)}\log\frac{\Lambda^2}{M_V^2} \right. \nonumber \\
 & &  + \rho_{A}\rho_{V'} \frac{\sigma-\rho_{V'}}{(\rho_{V'}-1)(\rho_{V'}-\rho_{A})}
 \log\frac{\Lambda^2}{M_V^2 \rho_{V'}} \nonumber \\
 & &  \left. -\rho_A^2\frac{\sigma-\rho_{A}}{(\rho_{A}-1)(\rho_{V'}-\rho_{A})}\log\frac{\Lambda^2}{M_V^2 \rho_A}\right]\,. \label{O1} 
\eea}

\noi
For $\rho_{V'}=\sigma$ this expression reduces to the MHA discussed in ref.~\cite{KPdeR01}.

\vspace*{0.25cm}
\item[Evaluation of $\langle O_2 \rangle$ ]

This is a vev which appears in the short--distance behaviour of the $\Pi_{LR}(Q^2)$ function; more precisely
\begin{equation}
\lim_{Q^2\ra\infty}\left(-Q^2\Pi_{LR}(Q^2)\right)\times Q^4=4\pi^2\frac{\als}{\pi}\left( 4\langle O_2\rangle +\frac{2}{N_c}\langle O_1 \rangle\right)+\cO\left(\frac{\als}{\pi} \right)\,.	
\end{equation}
It is related to the evaluation of matrix elements of the $Q_8$ operator as follows
\be
\lbl{Q8}
M_8 \equiv
\langle(\pi\pi)_{I=2}\vert Q_8\vert K^0 \rangle = \frac{8}{F_0^3}\langle O_2\rangle\,.
\ee
As discussed in ref.~\cite{KPdeR01}, to a good approximation and including next--to--leading $\als$--corrections we have
\be
\langle O_2(\mu)\rangle \simeq  \frac{1}{16\pi\als(\mu)}\left(\sum_A f_A^2 M_A^6 - \sum_V f_V^2 M_V^6 \right)\times\left[
1-\left( \begin{array}{c}25/8\ ({\mbox{\rm\small  NDR }})\\  21/8\ ({\mbox{\rm\small HV }})
\end{array}\right)\frac{\als{ (\mu)}}{\pi}\right]\,;
\ee
and in the case of a $\pi$--$V$--$A$--$V'$ spectrum

{\setl
\bea
 \langle O_2(\mu)\rangle & = & \frac{ M_V^6 }{16\pi\als(\mu)} \rho_{F}
 \frac{\rho_{V'}}{\sigma}
 \left\{\rho_A^2 \frac{\sigma-\rho_{A}}{(\rho_{A}-1)(\rho_{V'}-\rho_{A})}\right. \nn \\
  & & \left. -\rho_A
 \frac{\sigma-1}{(\rho_A-1)(\rho_{V'}-1)} -\rho_A\rho_{V'}
 \frac{\sigma-\rho_{V'}}{(\rho_{V'}-1)(\rho_{V'}-\rho_{A})} \right\}\times \nn \\
 & & \left[
1-\left( \begin{array}{c}25/8\ ({\mbox{\rm\small  NDR }})\\  21/8\ ({\mbox{\rm\small HV }})
\end{array}\right)\frac{\als(\mu)}{\pi}\right]\,, \label{O2}
\eea}

\noi
Again, in the limit where $\rho_{V'}=\sigma$ this expression reduces to the corresponding MHA expression discussed in ref.~\cite{KPdeR01}.

\end{description}

\vspace*{0.25cm}
\item{\sc Duality Constraint}
  
This can be formulated 	as the requirement that in the chiral limit, there is no $1/Q^2$--term in the OPE of the Adler function defined as
\begin{equation}
	\cA(Q^2)=\int_0^{\infty}dt\frac{Q^2}{(Q^2+t)^2}\frac{1}{\pi}\Imm\Pi_V(t)\,.
\end{equation}
The Adler function is not an {\it order parameter} of S$\chi$SB and, therefore, it has contributions from the perturbative continuum. Then, in the case of two explicit $V$ and $V'$ states plus a continuum spectrum, the requirement in question reads as follows
\be
2f_V^2 M_V^2+2f_{V'}^2 M_{V'}^2=\frac{N_c}{16\pi^2}\frac{4}{3}s_0 \left(1+\frac{3}{8}\frac{N_c\als(s_0)}{\pi}+\cdots \right)\,.
\ee
Using the 1st Weinberg sum rule, this can be written as a simple constraint between $\rho_F$, $\rho_A$, $\rho_{V'}$, $\sigma$ and the onset of the continuum $s_0$ which, obviously, has to start beyond the $\rho_{V'}$--pole; i.e., $s_0 > M_{\rho}^2 \rho_{V'}$: 
\be
\lbl{sth}
\rho_F\left(1+\frac{\rho_{V'}}{\sigma}\frac{\sigma -\rho_A}{(\rho_A -1)(\rho_{V'}-\rho_A)} \right)=
\frac{N_c}{16\pi^2}\frac{2}{3}\frac{s_0}
{M_V^2} \left(1+\frac{3}{8}\frac{N_c\als(s_0)}{\pi}+\cdots \right)\,.
\ee
		
\end{enumerate}

\section{Numerical Analyses and Conclusions}
\setcounter{equation}{0}

\begin{enumerate}
\item {\sc The case of a $\pi - V - A$ spectrum (MHA)}

(a) \textbf{Fixing the free parameters $\rho_F$ and $\rho_A$. }

Confronting the MHA approximation to the experimental values of the observables introduced in the previous section will allow us to test its consistency and adjust the two free parameters $\rho_F$ and $\rho_A$ of this approximation. 
We use as input the following set of experimental data 
\begin{align}
&\label{deltampi}\delta m_\pi = 4.5936 \pm 0.0005 \;\MeV\,, &{\mbox{\rm ref.}}~\cite{PDG03} \\
&L_{10} = (-5.13 \pm 0.19)\times10^{-3}\,, &{\mbox{\rm ref.}}~\cite{DGHS98} \\
&\Gamma_{\rho \ra e^+e^-} = (6.77 \pm 0.32)\times10^{-3} \;\MeV\,, &{\mbox{\rm ref.}}~\cite{PDG03}\\
&\Gamma_{a \ra \pi\gamma} = (640 \pm 246)\times10^{-3} \;\MeV\,, &{\mbox{\rm ref.}}~\cite{PDG03}\\
&\label{L9}L_9 = (6.9 \pm 0.7) \times 10^{-3}\,,&{\mbox{\rm ref.}}~\cite{BEG94}\\
& M_{\rho}= (775.9 \pm 0.5)\;\MeV\,; &{\mbox{\rm ref.}}~\cite{PDG03}
\end{align}

and make the reasonable assumption that these observables follow gaussian probability density functions (p.d.f.). 
In fact, some of these observables, when expressed in terms of the MHA parameters depend not only on $\rho_F$ and $\rho_A$ but also on $(m_{\pi^+} + m_{\pi^0})$  and/or $M_V$. Therefore, it is more appropriate for our purposes to use in our fit procedure the dimensionless quantities: 
\begin{align}
&\label{ratio1}\frac{m_{\pi^+} + m_{\pi^0}}{M_\rho^2}\delta m_\pi = \frac{3}{4}\frac{\alpha}{\pi} \frac{\rho_A \log (\rho_A)}{\rho_A-1}\,,\\
&L_{10} =- \frac{1}{4} \rho_F \left( 1 + \frac{1}{\rho_A}\right)\,,\\
&\label{ratio2}\frac{1}{M_\rho}\Gamma_{\rho \ra e^+e^-} = \frac{4\pi\alpha^2}{3}\frac{\rho_A}{\rho_A-1}\,,\\
&\label{ratio3}\frac{1}{M_\rho}\Gamma_{a \ra \pi\gamma} = \frac{\alpha}{24} \frac{\sqrt{\rho_A}}{\rho_A-1}\,,\\
&\label{L9}L_9 = \frac{1}{2}\rho_F\;.
\end{align}

This, however, has the drawback that the three ratios (\ref{ratio1}), (\ref{ratio2}) and (\ref{ratio3}) may  no longer have simple gaussian p.d.f. In order to check this, we have computed their p.d.f.~\footnote{We do this following the example given in ref.~\cite{HLLD01}.} and found that for each of them, there is practically no numerical difference between the  calculated p.d.f. and the gaussian one. This justifies the use of the standard $\chi^2$ statistical regression method to fit our parameters $\rho_F$ and $\rho_A$ with the result
\begin{equation}
 \rho_F=(12.36 \pm 0.35)\times 10^{-3}\quad \annd \quad \rho_A=1.464 \pm 0.004\,,	
\end{equation}
with a $\chi^2_{\text{min}}=1.21$ for 3 degrees of freedom (dof). The covariance matrix is given by
\begin{equation}
\text{cov} \left(\rho_F, \rho_A\right) = 
\begin{pmatrix}
1.21 & 1.36 \\
1.36 & 162 
\end{pmatrix}\times 10^{-7}\;.
\end{equation}

The quoted statistical errors of $\rho_A$ and $\rho_F$ have been obtained by evaluating the $1\sigma$ standard deviation via the solutions of:  $\Delta \chi^2(\rho_i) \doteq \min_{j,j\neq i}\left(\chi^2 (\rho_i,\rho_j) - \chi^2_{\text{min.}}\right)= 1$. 

We deduce from these results two conclusions: first that the MHA framework is statistically relevant and second that the fitted free parameters have small statistical errors. Moreover, they obey a multivariate gaussian p.d.f.. For $M_V=(775.9 \pm 0.5)\;\MeV$, we find $F_0=(86.3 \pm 1.2)\; \MeV$ and $M_A= (938.7 \pm 1.4)\;\MeV$, where the errors are only the statistical errors of the fit. The corresponding perturbative threshold $s_0$ defined by Eq.~\rf{sth} is $\sqrt{s_0} \approx 1.3 \;\GeV$.

(b) \textbf{Predictions of the MHA}

Using Eq.~(\ref{OnMHA}), we have evaluated the first few condensates of lowest dimension. The quoted numbers are given in Table~1 below in the entry MHA (the second line). They have been obtained using a Monte-Carlo simulation which takes into account the statistical correlation between the two parameters. The second error is an estimate of the systematic  theoretical error which we have made in the following way: 

i) We give a {\it systematic error} to the parameter $\rho_F$ of $\cO(\Gamma_{\rho}/M_{\rho})$. Notice that this is the parameter which modulates the large--$N_c$ counting in all our theoretical expressions (both for the MHA and the MHA + V').

ii) We enlarge the experimental error of $M_{\rho}$, which is the quantity modulating the dimensions of the calculated observables, by a factor of twenty i.e.,
\begin{equation}
	M_{\rho}\ra 	M_{\rho}=(776\pm 10)~\MeV\,.
\end{equation}

iii) As already mentioned in the text, the MHA + V' framework reduces to the MHA one for $\sigma = \rho_{V'}$. This suggests a way of introducing an extra {\it systematic error} to the MHA results versus the MHA plus an extra pole (R) results, by fixing the {\it a priori} ignorance one has on the relative position of an extra pole versus an extra zero within reasonable limits. We propose to quantify this error as follows 
\begin{equation}
\sigma=\rho_{R}\pm\frac{\rho_{R}-\rho_{A}}{2}\,,
\end{equation}
using the experimental values for $\rho_{A}$ and $\rho_{R}$, $(\rho_{A}=2.5$ and $\rho_{R}=3.6)$. Notice that this covers the possibility that the extra pole is of the $V$--type $(\sigma>\rho_R)$ or of the $A$--type $(\sigma<\rho_R)$.

The three sources of {\it systematic errors} are then added in quadrature.

For the purposes of comparison we also show in the same Table~1 the results of the other determinations of the chiral condensates; in particular the values quoted by  Cirigliano {\it et al.} in ref.~\cite{CDG00}~\footnote{See the original reference for a discussion of the two sources of errors.}. Notice that these authors also find an alternance of signs, but opposite to our MHA prediction, except for the lowest dimension condensate.

\begin{table*}[h]
\caption[Results]{\sc Numerical Results for the Chiral Condensates}
\lbl{table1}
\vspace*{0.5cm}
\hspace*{-2.4cm}
\begin{tabular} [c] {|c|c|c|c|c|c|c|}
\hline\hline
 & $\left< \mathcal{O}_6\right>$ & $\left< \mathcal{O}_8\right>$ & $\left< \mathcal{O}_{10}\right>$ & $\left< \mathcal{O}_{12}\right>$ & $\left< \mathcal{O}_{14}\right>$ & $\left< \mathcal{O}_{16}\right>$ \\
 & $\times 10^3 \; \; \text{GeV}^6$ & $\times 10^3 \; \; \text{GeV}^8$ & $\times 10^3 \; \; \text{GeV}^{10}$ & $\times 10^3 \; \; \text{GeV}^{12}$ & $\times 10^3 \; \; \text{GeV}^{14}$ & $\times 10^3 \; \; \text{GeV}^{16}$ \\
\hline\hline
 & & & & & & \\
MHA + V' & $-7.90 \pm 0.20 $ & $ +11.69 \pm 0.32$ & $- 13.12 \pm 0.43$  & $+ 13.21 \pm 0.62$ & $- 12.54 \pm 0.93$ & $+ 11.45 \pm 1.50$\\
 & $\pm1.62$ & $\pm2.53$ & $\pm3.01$  & $\pm3.24$ & $\pm3.29$ & $\pm3.21$\\
\hline 
 & & & & & & \\
MHA & $-7.89\pm 0.23 $ & $+11.71 \pm 0.34$ & $-13.18 \pm 0.41$  & $+13.33 \pm0.42 $ & $-12.78 \pm0.43 $ & $+11.89 \pm0.40 $\\
 & $\pm2.01$ & $\pm3.08$ & $\pm3.61$  & $\pm3.83$ & $\pm3.86$ & $\pm3.78$\\
\hline
\hline  
ALEPH & $-7.7 \pm 0.8 $ & $ +11 \pm 1 $ &       &                  &                 & \\
\hline 
OPAL  & $-6.0 \pm 0.6 $ & $ +7.5 \pm 1.3 $ &       &                  &                 & \\
\hline 
Davier \textit{et al.} \cite{DGHS98} & $-6.4 \pm 1.6 $ & $ +8.7 \pm 2.4 $ &       &                  &                 & \\
\hline 
Ioffe \textit{et al.}\cite{IZ01}& $-6.8 \pm 2.1 $ & $ +7 \pm 4 $ &       &                  &                 & \\
\hline 
Zyablyuk \cite{Z04}& $-7.1 \pm 1.5 $ & $ +7.8 \pm 3.0 $ & $-4.5 \pm 3.4$   &                  &                 & \\
\hline
\hline 
Bijnens \textit{et al.} \cite{BGP01}& $-3.2 \pm 2.0 $ & $ -12.4 \pm 9.0 $ &            &                  &                 & \\
\hline 
Cirigliano \textit{et al.} & $-4.5 \pm 0.83$ & $-5.7 \pm 3.72 $ & $+48.2 \pm 10.2$&$-160 \pm 26 $&$+426 \pm 62$ & $-1030 \pm 140$ \\
ALEPH\cite{CGM03}& $\pm 0.18$ & $\pm 0.64$ & $\pm 2$&$\pm 5$&$\pm 14$& $\pm 30$ \\
\hline
Cirigliano \textit{et al.} & $-5.06 \pm 0.89$ & $ -3.12 \pm 3.82 $ & $+38.7 \pm 10.6$&$-132 \pm 27 $&$+354 \pm 66$ & $-850 \pm 150$ \\
OPAL\cite{CGM03}& $\pm 0.12$ & $\pm 0.45$ & $\pm 1$&$\pm 3$&$\pm 6$& $\pm 20$ \\
\hline
Rojo \textit{et al.}\cite{RL04} & $-4 \pm 2 $ & \small$-12 ^{+7}_{-11} $ & $ 78 \pm 24 
$ & $-260 \pm 80
$&          &\\
\hline\hline
\end{tabular}
\end{table*}

\noi
Using Eqs.~\rf{Q7} and \rf{Q8} we can also evaluate the matrix elements of $Q_7$ and $Q_8$ operators. The results are given in the line MHA of Table~2 below. Again, the first error is statistical, the second error is our estimate of the theoretical systematic error in the way described above. 

\begin{table*}
\caption[Rs]{\sc Numerical Results for the $Q_7$ and $Q_8$ Matrix Elements}
\lbl{table2}
\vspace*{0.5cm}
\hspace*{-1.8cm}
\begin{tabular} [c] {|c|c|c||c|c|}
\hline\hline
 & $M_7$ (NDR) &  $M_7$ (HV) & $M_8$ (NDR) &  $M_8$ (HV) \\
\hline
MHA + V' & $0.12 \pm 0.00 \pm 0.01 $ & $0.59 \pm0.01 \pm0.06$ & $2.00 \pm 0.03 \pm 0.20$ & $2.15 \pm 0.03 \pm0.22$ \\
\hline 
MHA & $0.12 \pm 0.00 \pm 0.02$ & $0.59 \pm 0.01 \pm 0.11$ & $1.99\pm 0.03 \pm 0.36$ & $2.15 \pm 0.03 \pm 0.39$ \\
\hline
\hline
Donoghue \textit{et al.} \cite{CDGM01} & $0.16 \pm 0.1$ & $0.49 \pm 0.07$ & $2.22 \pm 0.67$ & $2.46 \pm 0.70$\\
\hline
Bijnens \textit{et al.} \cite{BGP01} & $0.24 \pm 0.03$ & $0.37 \pm 0.08$ & $1.2 \pm 0.8$ & $1.3 \pm 0.6$\\
\hline
Cirigliano \textit{et al.} \cite{CDGM03}& $0.22 \pm 0.05$ & & $1.50 \pm 0.27$ & \\
\hline
Narison \cite{N01}& $0.21 \pm 0.05$ & & $1.4 \pm 0.35$ & \\
\hline
\hline
RBC \cite{RBC01} & $0.27 \pm 0.03 $ &  & $ 1.1 \pm 0.2 $ & \\
\hline
CP-PACS \textit{et al.} \cite{CP01}& $0.24 \pm 0.03$ & & $1.0 \pm 0.2$ & \\
\hline
Donini \textit{et al.} \cite{DGGM99}& $0.11 \pm 0.04$ & $0.18 \pm 0.06$ & $0.51 \pm 0.10$ & $0.62 \pm 0.12$ \\
\hline
Bhattacharia \textit{et al.} \cite{BHA02}& $0.32 \pm 0.06$ & & $1.2 \pm 0.2$ & \\
\hline
SPQCDR \cite{SPQCDR02}& $0.24 \pm 0.02$ & & $1.05 \pm 0.10$ & \\
\hline\hline
\end{tabular}
\end{table*}

The results in Table~1 and Table~2, corresponding to the MHA, are perfectly consistent with those previously obtained in refs.~\cite{PPdeR01} and \cite{PPdeR98} using a different treatment of the input parameters.

\item {\sc The case of a $\pi - V - A - V'$ spectrum (MHA+V')}

\vspace*{0.25cm}
(a) \textbf{Fixing the free parameters}

Adding a vector resonance $V'$ in the spectrum extends the number of free parameters from two to four: $\rho_F$, $\rho_A$, $\rho_{V'}$ and $\sigma$. Furthermore, Eq.~(\ref{L9}) becomes now a function of $g_V$ as explained in Section~III.7. The way we treat this is by considering the observable\footnote{The decay width of $\rho \rightarrow \pi\pi$ is $\Gamma_{\rho \rightarrow \pi\pi}=(150.4\pm 1.3) \MeV$ \cite{PDG03}.}:
\begin{equation}
\frac{1}{M_\rho}\Gamma_{\rho\rightarrow \pi\pi} = \frac{1}{48\pi} \frac{1}{\rho_F^2}\frac{\sigma \left(\rho_F - 2 L_9 \rho_{V'}\right)^2(\rho_A-1)(\rho_{V'}-1)}{(1-\rho_{V'})^2\rho_F\rho_A \rho_{V'}(\sigma-1)}\;,
\end{equation}
as a function of $L_9$ which has an error itself, and is added as an extra parameter in our fit. The number of d.o.f. does not change since $L_9$ is also taken as an observable. 
We also impose a criterion of rejection through the ordering: 
\begin{equation}
\rho_A<\rho_{V'}<\sigma < \rho_0 \doteq \frac{s_0}{M_V^2} \;.	
\end{equation} 
The first and second inequalities reflect the knowledge that the new state has a higher mass than the axial, that it is a $V$--like pole and, therefore, its residue contributes positively to the $W_{LR}(z)$--function; the third inequality follows from the requirement  that the perturbative threshold $s_0$ defined in Eq.~\rf{sth} already lies  beyond the radius where the analytic structure of the poles and zeros retained satisfies the leading OPE constraint.

A similar statistical analysis to the one in the previous subsection with a $\chi^2$ regression leads to the following results:
{\setl
\bea
\rho_F & = & (12.36 \pm 0.03)\times 10^{-3}\,, \lbl{rhoF} \\
\rho_A & = & 1.466 \pm 0.003\,, \\
\rho_{V'} & =  & 2.63 \pm 0.01\,, \\
\sigma & = & 2.64 \pm 0.01\,. \lbl{sigma}
\eea}

\noi
The results in Eqs.~\rf{rhoF} to \rf{sigma} correspond to a value:  $L_9=(6.44 \pm 0.02)\times 10^{-3}$, with a $\chi^2_{\text{min}}=0.60$ for $1$ dof. The errors, which are only the statistical errors of the fit, were calculated using the same ``reduced-$\chi^2$'' procedure as before.   
We find that the parameters $\rho_F$ and $\rho_A$ are statistically stable when compared to those found in the MHA case.
We also find that $\rho_{V'} \approx \sigma$, which is consistent with the fact that the MHA approximation seems to have already the bulk of the full large--$N_c$ information. In other words, adding an extra  V'--pole appears to be compensated, at a very good approximation,  by the position of the nearby zero. 

In order to make numerical predictions for the chiral condensates and the matrix elements $M_7$ and $M_8$  we need to know the shape of each "reduced-$\chi^2$" so as to implement a  Monte-Carlo simulation. Concerning $\rho_F$ and $L_9$ we find parabolic shapes, i.e.  gaussian behaviours. In the other cases the structure of the "reduced-$\chi^2$" are slightly more complicated but since, as already seen in the MHA case, the resulting statistical errors remain very small as compared to the systematic  errors of the theoretical framework, we have finally assumed that all our parameters are gaussian and uncorrelated.
For $M_V=(775.9 \pm 0.5)\;\MeV$, we now find $F_0= (86.1 \pm 1.1)\; \MeV$, $M_A= (939.4 \pm 1.1 )\;\MeV$ and $M_{V'}=(1258.2\pm 2.5)\;\MeV$, where the errors are only the statistical errors of the fit, and a  perturbative threshold in the $V$--channel (or $A$--channel) starting, again, at $\sqrt{s_0} \approx 1.3 \;\GeV$.

\vspace*{0.25cm}
(b) \textbf{Predictions of the MHA+V'}

The restriction of (\ref{condensatgene}) to the $P=3$ case (MHA+V') reads 
\begin{equation}
\left < \cO_{2n+4} \right> = 2 (-1)^{n+1}\, M_V^{2n+4}\, \rho_F \sum_{k=1}^3 \left(1-\frac{\rho_k}{\sigma}\right) \prod_{i= 1}^{3} \frac{\rho_i}{\left(\rho_i - \rho_k + \delta_{ik}\right)} \rho_k^n \;,
\end{equation}
where $\rho_1=1$, $\rho_2=\rho_A$ and $\rho_3=\rho_{V'}$. The resulting values of the chiral condensates are given in the first line MHA + V' in Table~1.
The matrix elements of $Q_7$ and $Q_8$ are now obtained using Eqs.~\rf{Q7}, (\ref{O1}) and \rf{Q8},(\ref{O2}) and the corresponding results are given in the first line MHA + V' of Table~2. The systematic errors of the MHA + V' predictions have been estimated using the prescriptions i) and ii) already described earlier for the MHA predictions. Within errors, the two set of predictions from MHA and from MHA + V' are perfectly consistent with each other. 

\begin{figure}[h]

\begin{center}
\includegraphics[width=0.6\textwidth]{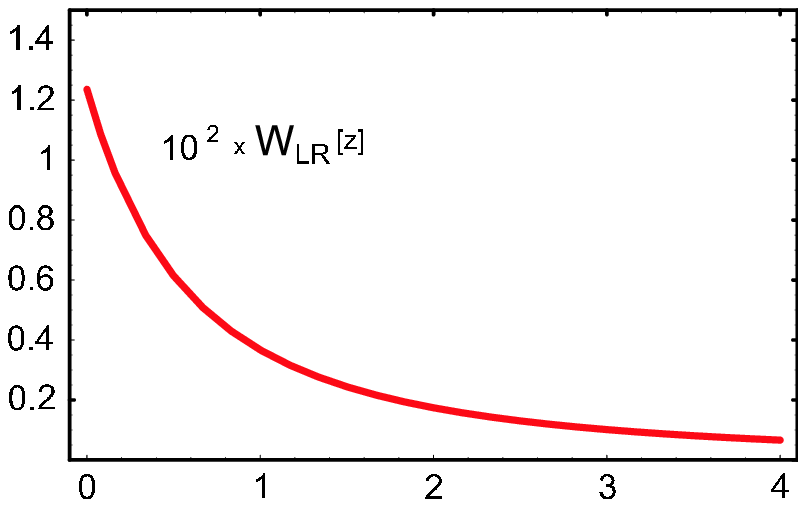}

{\bf Fig.~2a} {\it The predicted shape of the function $W_{LR}[z]$ in Eq.~\rf{LRV2} in the euclidean.}
\end{center}

\end{figure}

The shape of the function $W_{LR}[z]$ in Eq.~\rf{LRV2} with the parameters fixed at the central values resulting from the fit in Eqs.~\rf{rhoF} to~\rf{sigma} is shown in Figs.~2a,~2b and 2c. Figure 2a shows the shape of the function  $W_{LR}[z]$ in the euclidean region $(z\ge 0)$. Figures 2b and 2c show the shape of $\Ree W_{LR}[z]$ (the thick solid red lines) in the Minkowski region for $-2\le z\le 0$ in Fig.~2b and in the region of the $V'$ in Fig.~2c. Notice the different scales of the three figures. The delta functions contributing to $\Imm W_{LR}[z]$ are also shown (the thin vertical blue lines $V$ and $A$ in Fig.~2b, and the thin vertical blue line $V'$ in Fig.~2c.)

\begin{figure}[h]

\begin{center}
\includegraphics[width=0.6\textwidth]{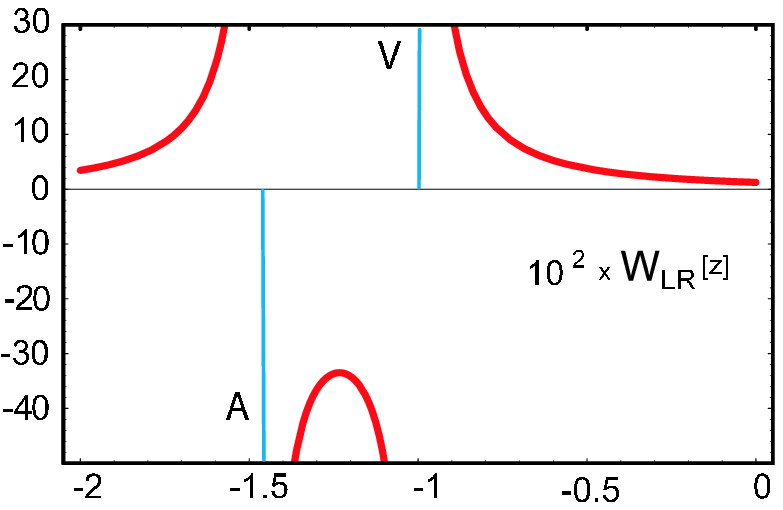}
\end{center}

{\bf Fig.~2b} {\it The predicted shape of the function $W_{LR}[z]$ in Eq.~\rf{LRV2} in the Minkowski region $-2\le z\le 0$. The vertical lines are the delta functions contributing to the imaginary part, the continuous solid lines, are the shape of $\Ree W_{LR}[z]$ in that region.}

\end{figure}

\begin{figure}[h]

\begin{center}
\includegraphics[width=0.6\textwidth]{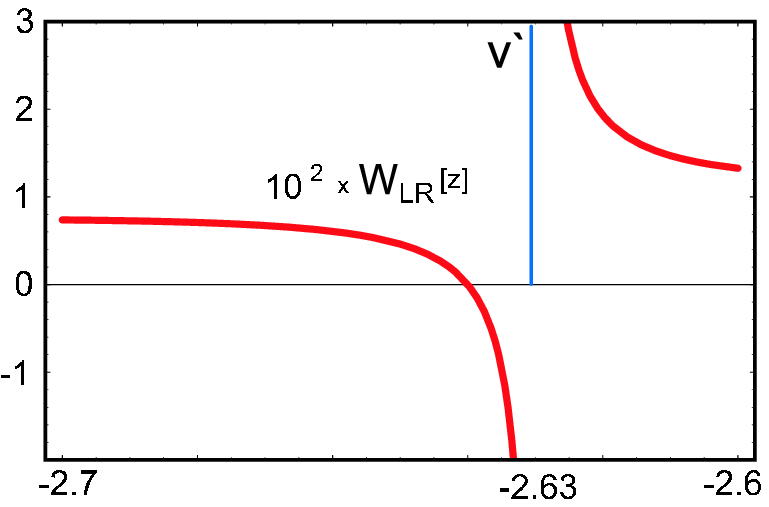}

{\bf Fig.~2c} {\it The predicted shape of the function $W_{LR}[z]$ in the $V'$ region.}

\end{center}

\end{figure}

\item {\sc Comparison with other Determinations}

The authors of refs.~\cite{BGP01} and~\cite{CGM03} give results for quite a few observables. This offers the possibility of making a comparative study with our predictions. The way we do that is by asking the following question:  {\it ``what are the values of $\rho_F$, $\rho_A$, $\rho_{V'}$ and $\sigma$ which, using the large--$N_c$ parameterization MHA +V' given in the text,  can fit the predictions of these authors?''} 

Concerning the work of  Bijnens \textit{et al.}~\cite{BGP01}, we have used their predictions (with their errors) for $\left< \cO _6\right>$, $\left< \cO _8\right>$, $M_7$, $M_8$ and the perturbative threshold (which in the case of  Aleph they take at $s_0=2.53^{+0.13}_{-0.12}\;\GeV$) as input data. Using this input, and imposing the constraint $\rho_A < \rho_{V'} < \sigma <\rho_0$ it is possible to find a fit with a $\chi^2_\text{min.}= 11.9 / (3 \;\text{dof}) $ and the central values:

\be
\rho_F = 0.011_,,\quad
\rho_A = 1.69\,,\quad
\rho_{V'}=2.08\,,\quad
\sigma = 4.10\,,
\ee
with
\be
F_0 = 81.2 \; \text{MeV}\quad\annd\quad
\rho_{0}=\frac{\rho_F}{F_0^2}s_0= 4.34\,.
\ee
The problem with this fit, which reflects the rather bad $\chi^2_\text{min.}$, is that $\sigma < 1+\rho_A +\rho_{V'}$ in contradiction with the equal sign value for $\cO_6$ and $\cO_8$ which these authors find. 

It is possible, however, to find a reasonable fit, using as input the predictions of  Bijnens \textit{et al.}~\cite{BGP01}, if one leaves free the position of $\sigma$. We have found a solution with a $\chi^2_\text{min.}= 1.2 / (3 \;\text{dof}) $ and parameter values:

\begin{align}
&\rho_F = 0.0147 ^{+0.0010}_{-0.0008} \\
&\rho_A = 3.4 ^{+1.8}_{-1.5}\\
&\rho_{V'}=3.7 ^{+1.0}_{-1.2} \\ 
&\sigma = 15.55 ^{+5.75}_{-3.25}\\
&F_0 = 86.9 ^{+5.3}_{-5.5}\; \text{MeV}\\
&\rho_{0}= 5.18\,,
\end{align}

where the errors are only the statistical errors of the fit. The problem of this fit, however, is that the position of the zero is far beyond the onset of the pQCD threshold $(\sigma \gg \rho_{0})$ and many states, which have not been included in the analysis, can fill this gap.

Concerning the work of Cirigliano \textit{et al.}~\cite{CGM03}, we have used their values for the condensates from $\cO_6$  to  $\cO_{16}$, with their errors, as well as their  predictions of $M_7$ and $M_8$, with their errors as well, as input values. Like in our previous analysis, we have first imposed the constraint $\rho_A < \rho_{V'} < \rho_0^\text{sup.}$ and $\sigma< \rho_0^\text{sup.}$ for $s_0=\{1.95 , 2.15, \cdots, 3.15\} \, \text{GeV}^2$. We have found a solution with, however, a very bad $\chi^2_\text{min.}= 134.5/ (4 \;\text{dof})$. The central values of the resulting parameters are 
\be
\rho_F = 0.010\,,\quad
\rho_A =  1.60\,,\quad
\rho_{V'}= 1.78\,,\quad 
\sigma =  3.48\,,
\ee
with
\be
F_0 =  93.7\; \text{MeV}\quad\annd\quad
\rho_0 = 3.59\,.  
\ee
Again, we find that this solution satisfies the relation $\sigma < 1+\rho_A +\rho_{V'}$ in contradiction with the equal sign value for $\cO_6$ and $\cO_8$ which the authors find.

If we relax the constraint on $\sigma<\rho_{0}$ it is then possible to find a ``better fit'' with a $\chi^2_\text{min.}= 3.26/ (4 \;\text{dof})$ and central values for the parameters:
\be
\rho_F =  0.009\,,\quad
\rho_A =  1.02\,,\quad
\rho_{V'}=  2.03\,,\quad
\sigma = 5.40\,,
\ee
with
\be
F_0 = 92.39\; \text{MeV}\quad\annd\quad
\rho_0= 3.47\,.
\ee 
The problem with this fit, however, is twofold. On the one hand the position of the axial state is too near to the first vector state $(\rho_{A}=1.02)$ and also the fact that there is still a large gap between the onset of the pQCD continuum $(\rho_0 =3.47)$ and the position of the zero $(\sigma=5.40)$.
	
We conclude from these analyses that the results of Bijnens \textit{et al.}~\cite{BGP01}, as well as the results of Cirigliano \textit{et al.}~\cite{CGM03}, if interpreted within the framework of large--$N_c$ QCD, show internal inconsistencies. As  recently discussed by Donoghue~\cite{DON04}, this could very well be the reflect of unphysical extrapolations in the hadronic spectral function which (implicitly or explicitly) have been made in the phenomenological analyses reported in refs.~\cite{BGP01,CGM03,RL04}.

\end{enumerate}
   
\vspace*{1cm}
\acknowledgments{We are very grateful to Santi Peris for many discussions on the topics reported here and to J\'er\^ome Charles for his generous help with the numerical analysis. This paper was finished while one of the authors (E~de R) was attending the Benasque Workshop on {\it ``Matching Light Quarks to Hadrons''}. It is a pleasure to thank the organizers for the very pleasant and stimulating environment they provided.}

\vspace*{0.5cm}

\noi
This work has been supported in part by TMR, EC-Contract No. HPRN-CT-2002-00311 (EURIDICE).

\end{document}